\documentclass[12pt,a4paper]{article}
\usepackage[width=17cm,centering]{geometry}

\usepackage{amssymb,amsmath}
\usepackage{graphicx}
\usepackage{dcolumn}
\usepackage{bm}
\usepackage{xcolor}
\usepackage{hyperref}
\usepackage{url}
\usepackage{cite}
\usepackage{float}

\usepackage{stmaryrd} 
\usepackage{wasysym} 

\usepackage[many]{tcolorbox}
\newtcolorbox{mybox}{
    enhanced, 
    breakable, 
    boxsep=7pt, 
    parbox=false, 
    before skip=15pt, 
    after skip=15pt,
    code={}
}

\newtcolorbox[blend into=figures]{myfigure}[2][]{
    enhanced, tile, flip title,
    float=htb,
    capture=minipage,
    fonttitle=\mdseries,
    coltitle=black,colbacktitle=black!5!white,
    before=\bigskip\begin{center},
    after=\end{center},
    title={#2},
    every float=\centering,
    #1
}

\newtcolorbox[blend into=tables]{mytable}[2][]{
    enhanced, size=small, tile, flip title,
    capture=minipage,
    fonttitle=\mdseries,
    coltitle=black, colbacktitle=black!5!white,
    halign title=flush center,
    before=\bigskip\begin{center},
    after=\end{center},
    title={#2},
    #1
}

\definecolor{forestgreen}{rgb}{0.33,0.61,0.34}

\title{Waiting-time paradox in 1922}
\bigskip
\author{Naoki Masuda${}^{1,2,*}$ and Takayuki Hiraoka${}^3$
\ \\
\ \\
${}^1$Department of Mathematics\\
University at Buffalo, State University of New York,\\
Buffalo, NY 14260-2900, USA\\
${}^2$Computational and Data-Enabled Science and Engineering Program,\\
University at Buffalo, State University of New York,\\
Buffalo, NY 14260-5030, USA\\
${}^3$Department of Computer Science,\\
Aalto University, Espoo 00076, Finland\\
\ \\
$^*$ Corresponding author (naokimas@buffalo.edu)
}

\begin{document}

\setlength{\baselineskip}{24pt}

\maketitle

\section*{Abstract}
We present an English translation and discussion of an essay that a Japanese physicist, Torahiko Terada, wrote in 1922. In the essay, he described the waiting-time paradox, also called the bus paradox, which is a known mathematical phenomenon in queuing theory, stochastic processes, and modern temporal network analysis. He also observed and analyzed data on Tokyo City trams to verify the relevance of the waiting-time paradox to busy passengers in Tokyo at the time. This essay seems to be one of the earliest documentations of the waiting-time paradox in a sufficiently scientific manner.

\section{Introduction}

Torahiko Terada (1878--1935) was a Japanese physicist and a professor at Tokyo Imperial University (which is now the University of Tokyo). After gaining his PhD in Tokyo, he studied in Berlin and Stockholm around 1910 as well as visiting other places before coming back to Japan. His work includes studies of X-ray diffraction~\cite{Terada1913Nature-1,Terada1913Nature-2,Terada1913ProcMathPhysSocTokyo}, which he published almost at the same time as William Henry Bragg and his son William Lawrence Bragg, who received the Nobel Prize in 1915. Apart from research, Terada also wrote various essays in plain Japanese, often describing his own analysis of everyday phenomena from a physicist point of view. Some of them may be regarded as a precursor of complexity science. Short stories about him are available in English~\cite{Matsushita2010EvolInstEconRev,Wittner2016book}.

In one of his essays, Terada documented what is known today in queuing theory, temporal network analysis, and other research fields as the waiting-time paradox, based on his empirical observations of Tokyo City tramway operation back in 1922. This essay, albeit written in Japanese, is probably one of the first work on the waiting-time paradox in a quantitative and practical context. It later inspired mathematical work on a Poisson process model for the congestion of trams~\cite{Nakatsuka1986JOperResSocJapan}.
In the following text, first, we briefly introduce the waiting-time paradox in Section~\ref{sec:waiting-time paradox}. Second, we provide an English translation of the entire essay in Section~\ref{sec:essay}\footnote{Another translation is available in a Japanese journal, of which the main purpose was to introduce Terada's work in both Japanese and English~\cite{Gally2013Kagaku-1,Gally2013Kagaku-2}.} (see Acknowledgments for information about the copyright). Third, we re-analyze his data in modern terms in Section~\ref{sec:modern analysis}. Finally, we conclude the article in Section~\ref{sec:conclusion}.

\section{Waiting-time paradox\label{sec:waiting-time paradox}}

The waiting-time paradox, also known as the inspection paradox and renewal-theory paradox, is a mathematical phenomenon known in the queuing theory and stochastic processes literature for a long time~\cite{Feller1971book2,Allen1990book,Masuda2016book}.
As its yet another alias, the bus paradox, suggests, we experience this paradox in everyday life when we are waiting at a bus stop for the next bus to arrive. Consider the time for which a passenger has to wait until the next bus comes since the passenger arrived at the bus stop, which is called the waiting time.
Here, we assume an inattentive passenger who arrives at the bus stop without checking the timetable, so that the time of arrival is distributed uniformly at random. 
The naive guess would be that the average waiting time is equal to half of the average time interval between two buses expected from the timetable. However, the waiting-time paradox states that it is in fact longer unless the buses arrive at the stop completely regularly.

Let us call the arrival of a bus an event.
The difference between the average waiting time until the next event and half of the average inter-event time is large if the inter-event time obeys a fat-tailed distribution. In various empirical data, from human activity to earthquakes, inter-event times often obey a fat-tailed distribution~\cite{Eckmann2004PNAS,Barabasi2005Nature,VazquezA2006PhysRevE-burst,GohBarabasi2008EPL,Barabasi2010book,HolmeSaramaki2012PhysRep,Karsai2018Springer}. Therefore, in these cases, which do not necessarily include the case of bus services, the waiting time can be much longer than half of the average inter-event time.
The waiting-time paradox elicits many interesting results on dynamics on temporal networks (i.e., time-varying networks), such as suppression of network-wide contagion when fat-tailed distributions of inter-event times are incorporated into epidemic process modeling~\cite{HolmeSaramaki2012PhysRep,Masuda2013F1000,Masuda2016book}.

\begin{figure}[tb]
\centering
\includegraphics[scale=0.55]{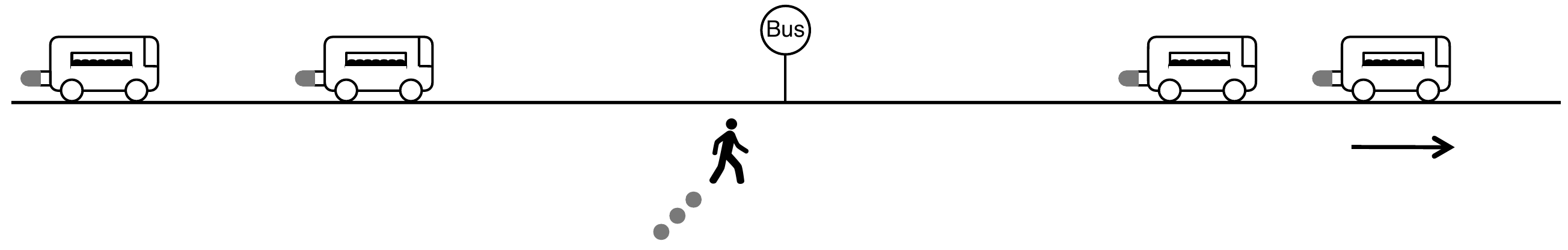}
\caption{Illustration of the waiting-time paradox. A uniformly randomly drawn time tends to fall within a long interval. 
}
\label{fig:waiting-time paradox}
\end{figure}

Mathematically, the paradox is formulated as follows. Suppose that the inter-event time, $\tau$, is independently drawn each time according to a probability density $\psi(\tau)$. Such a stochastic process is called a renewal process.
A Poisson process is a renewal process and defined by $\psi(\tau) = \lambda e^{-\lambda \tau}$ ($\lambda > 0$), i.e., an exponential distribution. 
Suppose that a passenger arrives at a bus stop at time $t_0$, which is uniformly distributed on the time axis.
Figure~\ref{fig:waiting-time paradox} schematically represents the situation. As the figure implies, $t_0$ is located inside an interval segmented by consecutive events with a probability proportional to the length of the interval. Therefore, the time $t_0$ tends to fall in a long interval rather than in a short one. This causes the waiting-time paradox.

The distribution of waiting times to the next event, $\psi^{\rm w}(t)$, where $t$ is the waiting time, is derived as follows. The probability density with which time $t_0$ belongs to an interval of length $\tau$ is given by
\begin{equation}
f(\tau) = \frac{\tau \psi(\tau)}{\int_0^{\infty} \tau^{\prime} \psi(\tau^{\prime}){\rm d}\tau^{\prime}} = \frac{\tau \psi(\tau)}{\langle\tau\rangle}.
\label{eq:fall within}
\end{equation}
Under the condition that $t_0$ belongs to an interval of length $\tau$, the probability density for the waiting time is given by
\begin{equation}
g(t|\tau) = \begin{cases}
\frac{1}{\tau} & (0\le t\le \tau),\\
0 & (t > \tau).
\end{cases}
\label{eq:waiting time pdf conditional}
\end{equation}
By combining Eqs.~\eqref{eq:fall within} and \eqref{eq:waiting time pdf conditional},
one obtains
\begin{equation}
\psi^{\rm w}(t) = \int_t^{\infty} f(\tau)g(t|\tau) {\rm d}\tau=
\frac{1}{{\langle\tau\rangle}} \int_t^\infty \psi(\tau) {\rm d}\tau.
\label{eq:waiting-time distribution}
\end{equation}
Using Eq.~\eqref{eq:waiting-time distribution}, one obtains the mean waiting time as follows:
\begin{align}
\int_0^{\infty} t \psi^{\rm w}(t) {\rm d}t
=& \frac{1}{\langle\tau\rangle}\int_0^{\infty} \left[t \int_t^{\infty} \psi(\tau) {\rm d}\tau\right] {\rm d}t\notag\\
=& \frac{1}{\langle\tau\rangle}\int_0^{\infty} \psi(\tau) \left[ \int_0^{\tau} t{\rm d}t\right] {\rm d}\tau\notag\\
=& \frac{\langle\tau^2\rangle}{2\langle\tau\rangle}.
\label{eq:mean waiting time}
\end{align}

Let us look at some examples.
First, when all inter-event times are equal to $\tau_0$,
one obtains $\langle\tau^2\rangle = \langle\tau\rangle^2 = \tau_0^2$. Equation~\eqref{eq:mean waiting time} implies that the mean waiting time in this case is given by $\langle\tau^2\rangle / \left(2\langle\tau\rangle\right) = \tau_0 /2$. Therefore, there is no ``paradox''.

Second, consider a Poisson process having $\psi(\tau) = \lambda e^{-\lambda \tau}$, 
where $\lambda = 1/\langle \tau\rangle$. One obtains
\begin{equation}
\psi^{\rm w}(t) = \frac{1}{{\langle\tau\rangle}} \int_t^\infty \lambda e^{-\lambda \tau} {\rm d}\tau
= \lambda e^{-\lambda t},
\end{equation}
such that the waiting-time distribution is the same as the inter-event time distribution, reflecting
the memoryless property of Poisson processes. In this case, because $\langle\tau^2\rangle = 2/\lambda^2$, the mean waiting time is equal to $\langle\tau^2\rangle / \left(2\langle\tau\rangle\right) = 1/\lambda = \langle\tau\rangle$, i.e., twice the naive guess (i.e., $\langle \tau\rangle / 2$). 
Note that the waiting-time paradox is in action for Poisson processes and renewal processes that have a small (but non-zero) variance in the inter-event times.

Our third example shows a more drastic and possibly more counter-intuitive effect of the waiting-time paradox. Consider a common form of power-law distribution of inter-event times given by
\begin{equation}
\psi(\tau) = \frac{\alpha-1}{(1+\tau)^{\alpha}}.
\label{eq:psi(tau) power law waiting-time paradox}
\end{equation}
With $\alpha > 2$, the mean inter-event time, $\langle \tau\rangle = (\alpha-2)^{-1}$,
is finite. The distribution of waiting times is given by
\begin{equation}
\psi^{\rm w}(t) = \frac{\alpha-2}{(1+t)^{\alpha-1}}.
\label{eq:waiting-time distribution PL1}
\end{equation}
Therefore, the tail of $\psi^{\rm w}(t)$ is fatter than that of $\psi(\tau)$, which suggests that the waiting times would be longer than the inter-event times on average. In fact, the mean waiting time is calculated as $\langle\tau^2\rangle / \left(2\langle\tau\rangle\right) = (\alpha-3)^{-1}$, assuming $\alpha > 3$, which is larger than the mean inter-event time, $\langle \tau\rangle = (\alpha-2)^{-1}$. 
When $\alpha \le 3$, the mean waiting time diverges because $\langle \tau^2 \rangle$ does, whereas the mean inter-event time remains finite if $\alpha>2$.
It should be noted that various empirical data such as human activity and earthquakes show power-law-like distributions of inter-event times~\cite{Eckmann2004PNAS,Barabasi2005Nature,VazquezA2006PhysRevE-burst,GohBarabasi2008EPL,Barabasi2010book,HolmeSaramaki2012PhysRep,Karsai2018Springer}.

\section{On congestion of trams\label{sec:essay}}

This section is devoted to a full translation of the Terada's essay~\cite{Terada1922essay} \footnote{
The original text in Japanese is freely available at \url{https://www.aozora.gr.jp/cards/000042/files/2449_11267.html}. The license available at \url{https://www.aozora.gr.jp/guide/kijyunn.html}, written only in Japanese, permits anybody to translate and modify it.}. See Refs.~\cite{Gally2013Kagaku-1,Gally2013Kagaku-2} for a different translation.
It was published in the September 1922 issue of Japanese journal Shiso (meaning ``Thought'')\footnote{This journal is still alive and has a reasonably good reputation.}. 
Here, we did not dare to try to convey his beautiful Japanese writing in English. We also avoided direct translation of the original sentences in some places to make reading smooth. The entire essay goes as follows.

\begin{mybox}
\setcounter{mpfootnote}{\value{footnote}}
For those who have weakened body and nerves, it is almost unbearable torture to be in a packed tram\footnote{Although the Japanese term used in the original text is ``trains", they were in fact street trams.}, clinging onto a strap, and being pushed around and stepped on by others. Even worse, the aftermath of traveling on a congested tram would last for hours. Since I recently got a chronic disease, I have been trying to travel on vacant services and to avoid congested ones.

My way of catching a vacant service is very simple: Just wait patiently until a vacant one comes. 

The most congested hours in a day seem to be fixed, depending on the line and direction. One may think that vacant services would not come in such rush hours even if one waits for long. However, if you keep waiting without a haste, you would sometimes encounter a non-congested service even in rush hours. This may sound strange, but it actually is not. This phenomenon has a due basis, which I noticed only recently. Until then, I had just recognized and taken advantage of this phenomenon as an empirical fact.

In the peak times, tram after tram is extremely full in a way that goes beyond the normal level of congestion\footnote{Note that he was Japanese. Those services being ``extremely full" to his eyes implies that they must have been insanely congested to non-Japanese eyes.}. Still, if I stay at the tram station and observe what happens for 10 to 15 minutes, I notice a constant rhythm in the level of congestion across different services. After observing six to seven services, I would recognize that the level of congestion oscillates and comes back to the previous level. 

Such an oscillation is the most evident when the level of congestion is intermediate, rather than when it is extremely high. If you stay at a station in such an intermediately congested hour, you will almost always find the following periodic pattern.

At first, you would see 10 to 20 people at the station. Most of them are looking at the direction from which trams come with a nervous look on their faces. The crowd size monotonically increases during this period. After five to seven minutes, a service finally arrives at the station. The people would then rush into the tram car, as if they cannot spare any time for passengers who are getting off, as if this is the last service and no tram would come after that. However, it is most likely that the next service or yet another one will arrive only within tens of seconds, or at most within two minutes. In the first congested tram, even the doorstep of the car is occupied by passengers. In contrast, in the second one, which arrives at the station almost when the first one departs, it is often the case that only one or two passengers are standing and holding on to straps (meaning all other passengers have seats).
In a lucky case, there are even some empty seats. In the third service, it is not uncommon that the car is almost empty after passengers have got off at the station.

After a couple of such empty services, no service would arrive for five to ten minutes. The number of passengers waiting at the station increases at a constant statistical rate during this period. The next service would arrive after the station has accumulated 20 to 30 people with already full of passengers. It discharges a couple of passengers and has to load 20 or 30 new passengers. A few waiting passengers who are unlucky enough to miss that service wait for another 30 seconds to end up catching the next service, in which they get to sit on a spacious seat with their legs stretched and enjoy a breeze coming in through opened windows if it is a hot day. When they arrive at their destination, they see that the service just ahead of theirs is still full. The passengers in the first service have to push people off like swimming in a muddy field to finally exit the tram, while those on the next service may get off at the same station almost at the same time.

I always avoid the ``peak" of this periodic dynamics of congestion and seek a ``valley". Then, I can secure a proper seat without any rush, feel calm, and do my reading\footnote{It is needless to say that this type is rare in contemporary Tokyo even after a century after his time.}. The time I have to waste while waiting for the valley of an oscillation, starting from its peak, is tens of seconds in a lucky case and will not exceed a minute or two anyways. While waiting, I would check items displayed in shops nearby, or research people's faces or morphology of clouds in the blue sky. I do waste a short amount of time due to this \footnote{Recall that he was a productive scientist. He must have been interested in efficiently managing his time, just like many of us are.}, but if it allows me to rest my body for tens of minutes on the tram and read ten pages of something which I would not otherwise read, I am pretty sure that my way is more beneficial. Furthermore, it is important for me to evade physical and mental fatigue that would overwhelm me after leaving a congested tram.

Anyways, I am hardly in a situation where arriving one or two minutes earlier to the destination is crucial. And I doubt how many of those people who are in hurry, unlike me, can be sure that they do not waste one or two minutes after leaving the tram.

All this is probably trivial. It is something that everybody should be fully aware of. Nevertheless, it seems that a great majority of passengers of Tokyo City trams think that they must catch the congested service that arrives right after a long interval of services. I do not understand this behavior, but this may be related to some strength of Japanese people's nature. For example, this may be correlated with the strength of Japanese in battlefields. Or, this may be a particular expression of something vague called modern thoughts. I should refrain myself from making easy critiques of this. What I want to consider here is not whether the majority's behavior is good or bad. Rather, I want to consider the scientific or mathematical problems of the rhythmic dynamics of congestion on trams, which inevitably stem from the behavioral tendency of passengers.

Consider the following case for simplicity.
Suppose that the services leave the terminal with a regular period, run at a constant speed, and are designed to stop at each station for a fixed amount of time. If this rule is strictly respected, the interval between the times that consecutive services pass an arbitrary fixed location on the route should be perfectly constant, which is denoted by $T$. However, in fact, because of unavoidable randomness arising from complex origins, each observed interval between services would be shifted to $T+\Delta T$. The deviation, $\Delta T$, may be large, small, positive, or negative and should distribute according to a Gaussian or similar distribution. In plain terms, some services arrive early and others arrive late, and different inter-service times would alternate in a complex manner. It should be noted that the average inter-service time is still equal to $T$. In other words, the sum of $\Delta T$ over different inter-service times is equal to 0.

Fluctuations in the time at which services arrive at a station can be easily calculated if the velocity of each service and the standard deviation of the time for which a service stops at a station are given. This being said, it is obvious that the fluctuations grow as the traveling distance increases. The fluctuations should be roughly proportional to the square root of the distance traveled.

The sequential order of different $\Delta T$'s (e.g., which value of $\Delta T$ comes first) is also governed by a law of randomness. This law is not an easy one, but in the end we observe that, on average, one in every three or four services is notably too early or too late.

The discussion so far ignores passengers' behavior. Let us now consider it.
The rate at which the number of passengers waiting at a station increases should have a mean, $n$, that depends on the time of the day and the location. The actual rate should fluctuate around $n$.
The mean number of passengers that a service loads at the station should be proportional to the time since the previous service left the station. If the previous service, $s_1$\footnote{Notations $s_1$ and $s_2$ are not in the original text, but we introduced them here with the aim of making the reading easier.}, leaves earlier than expected by time $a$ and the next service, $s_2$, arrives later than expected by time $b$, then service $s_2$ has to load $n(a+b)$ more people than the mean.
Skipping detailed calculations, it should be clear that, statistically speaking, services arriving late have to load more passengers.
Of course, we also need to consider passengers getting off the trams, but let us ignore them in this discussion.

We now discuss how the fluctuations in the number of passengers depending on services affect the duration for which a service stays at a station. If there are more passengers, the staying duration would be longer \footnote{This law certainly applies to the modern railway services in big cities in Japan.}. Even if all passengers are reserved gentlemen, the time necessary for them to get on board increases as the number of the passengers entering the tram increases. If these passengers try to coerce themselves in before the leaving passengers get off the tram, or if they argue with the conductor, the service has to stop at the station even longer. What are the consequences of the extended staying time?

Of course, it delays the arrival of service $s_2$ at the next station. This results in an increase in the number of additional passengers that it has to load at the next station, $nb$. Then, this service is further delayed and becomes even more congested. In a simplified case, the number of passengers to be loaded increases as a power of the number of stations at which the tram has stopped\footnote{Verifying this will need a careful analysis, which we do not attempt in this article.}. In reality, the number of passengers is limited by the capacity of the tram car, so it does not diverge. Anyways, the conclusion that congested services will get even more congested should be correct.

How about service $s_3$ that arrives just after this cursed service $s_2$? Suppose that
$s_3$ arrives at the first station on time. Because $s_2$ has been delayed by time $b$,
service $s_3$ has to load $nb$ less passengers than expected. Even if $s_3$ arrives late with delay
$c$, it still has to load $nb$ less passengers as compared to the case in which $s_2$ is not delayed\footnote{If $s_2$ is delayed by $b$ and $s_3$ is delayed by $c$, service $s_3$ has to load $n(b-c)$ less passengers as compared to when nothing has happened to either $s_2$ or $s_3$. In contrast, if $s_2$ is not delayed, $s_3$ has to load $-nc$ less passengers (i.e., $nc$ more passengers) than the mean, which is smaller than $n(b-c)$ by $nb$.}.
If $s_3$ arrives earlier than the schedule by time $c$, then it can depart with $n(b+c)$ less passengers than the mean.
What happens then? Obviously, the interval between $s_2$ and $s_3$ gradually decreases, and both the congestion of $s_2$ and the emptiness of $s_3$ will escalate\footnote{When the first author lived in Bristol, UK, it was more often than not the case that this interval was equal to 0 for local buses. It was not even uncommon that three buses bound for the same destination came in tandem. When he missed this wave, he had to wait three times more time than what the schedule said.}.

The tram cars that are initially evenly spaced on a long rail will have uneven inter-service times as the operation goes on. As a result, services that are delayed or advanced than the schedule would appear with a period of three or four services. Then, roughly speaking, what I described above happens, i.e., an extremely congested service would arrive every three or four services, and then the two or three services that follow will be gradually less congested. Then, after a long interval, the same thing will be repeated.

All this is a theoretical conclusion based on abstraction and deduction. It is needless to say that we should incorporate numerous other factors to further approach the reality. However, the insights obtained above are derived based on the most important factors among various ones. I believe that my conclusion is not too far from the reality.

To verify the theory, I occasionally measured the times at which actual services passed at given locations. As an example, the observations that I made on the evening of June 19, 1922, at a location 100 meters south of Jimbocho station, for services operating between Sugamo and Mita\footnote{This route was replaced by a subway line in 1973.}
are shown in Fig.~\ref{fig:jimbocho_observation}.
The symbols represent the levels of congestion that I roughly estimated (see the figure caption for details).

\begin{myfigure}[nofloat, label=fig:jimbocho_observation]{
Congestion of Tokyo City trams. 
(a) Original version in Japanese. 
(b) English translation. 
$\varocircle$: extremely congested, $\ocircle$: congested, $\triangle$: all seats are occupied but straps are largely free, $\times$: there are many empty seats, $\times\times$: there were only two or three passengers on board.
Panel (b) closely resembles an earlier English translation~\cite{Gally2013Kagaku-2}.}
    \centering
    \includegraphics[width=\columnwidth]{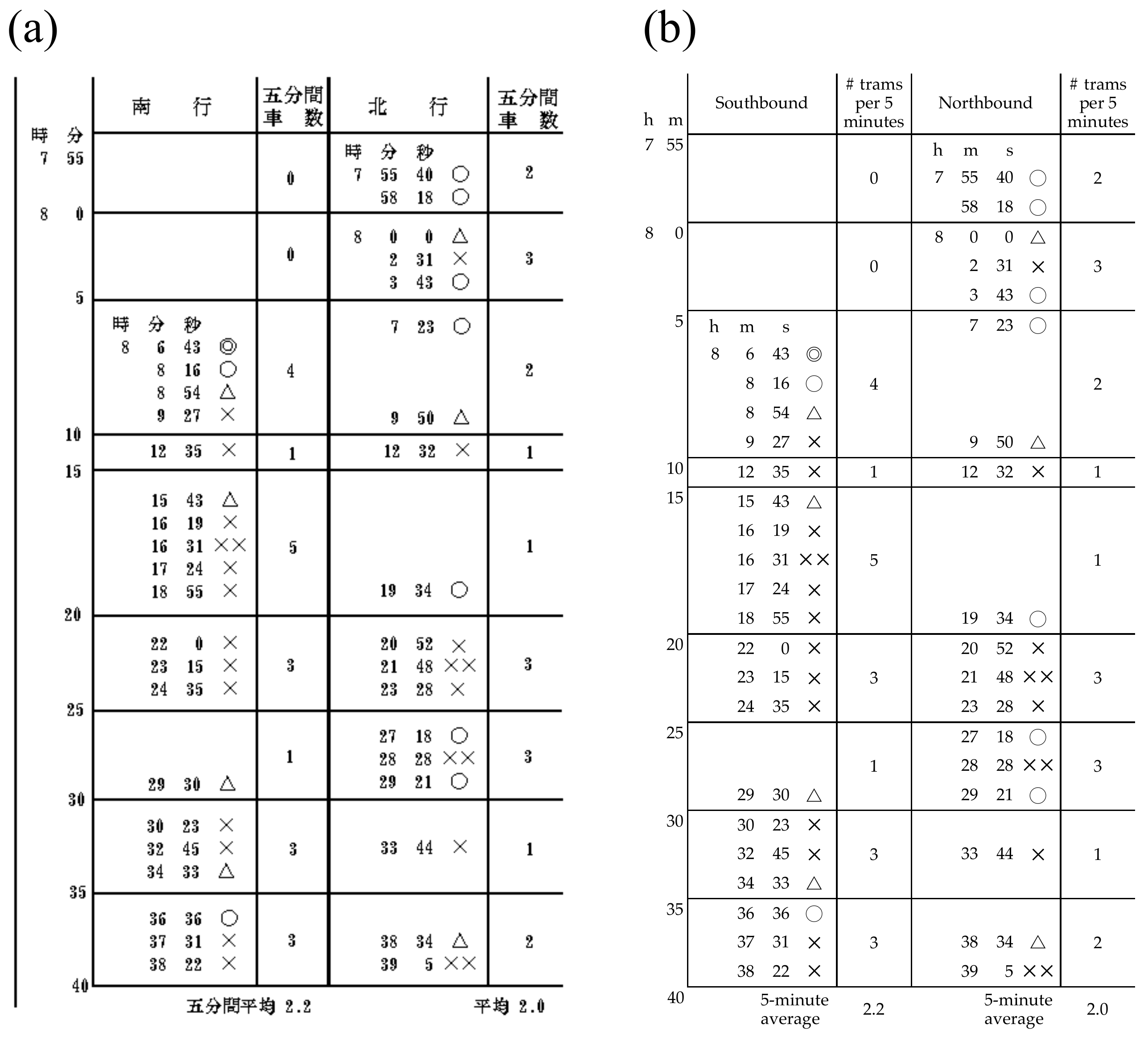}
\end{myfigure}

Figure~\ref{fig:jimbocho_observation} indicates the following. Although the number of services passing this location every five minutes is highly variable, the number averaged over the time is similar between southbound and northbound services; it is approximately one service per 2.5 minutes. However, the individual inter-service time varies from more than 11 minutes 3 seconds for the initial interval before the first southbound service in the observation to just 12 seconds. In addition, a long inter-service time tends to be followed by a congested service and vice versa.

\begin{mytable}[nofloat, label=tab:Terada_tau]{Distribution of inter-service times.}
\centering
\begin{tabular}{l|c}
inter-service time & frequency\\ 
\hline
$\ge$ 4 minutes & 4 \\
$\ge$ 3 minutes & 9 \\
$\ge$ 2 minutes & 15 \\
$\le$ 2 minutes & 23 \\
$\le$ 1 minute & 11 \\
$\le$ 40 seconds & 5
\end{tabular}\smallskip
\end{mytable}

Statistics of inter-service times is summarized in Table~\ref{tab:Terada_tau}\footnote{This table is not very comprehensible and contains error. It reads as follows. The inter-service time, denoted by $\tau$, is larger than or equal to four minutes four times. Because $\tau\ge 4$ implies $\tau\ge 3$, there are $9-4=5$ inter-service times with $3\le \tau < 4$. Likewise, there are $23-11=12$ inter-service times with $1 < \tau \le 2$. Because no inter-event time is exactly equal to two minutes, one does not have to worry about the overlap between the counts for $\tau\ge 2$ and $\tau\le 2$. Therefore, there is $15+23=38$ inter-service times in total, according to Terada's count. However, the table contains the following error. If we do not include the times from the beginning of the observation window (i.e., 7:55 pm) to the first southbound or northbound service in the sample of inter-service times, there are only 36 inter-service times in total. The row $\tau\le 2$ should have frequency 21, not 23. Even if we include these two durations and/or the times from the last service to the end of the observation window (i.e., 8:40 pm) as inter-service times such that the sample size becomes either 38 or 40, the distribution does not coincide with Table~\ref{tab:Terada_tau}.}.
The table indicates that there are more short inter-service times and fewer long inter-service times. Out of the 38 inter-service times, four (roughly 10\%) are longer than or equal to 4 minutes. However, we should not misunderstand the following thing: when it comes to the likelihood with which passengers encounter a certain inter-service time, it is a different story. To show this, one has to multiply the length of the inter-service time and the frequency of inter-service times of that length, sum this product over the range of time scales of interest, and compare the sum across different time scales\footnote{This is where he points out the waiting-time paradox.}. 
Let us classify the inter-service times with the window size of one minute, and calculate the sum of the products of the frequency of inter-service times falling within each class and the corresponding length of inter-service time.
Even if we neglect the inter-service times larger than five minutes, the sum of the product over the range of inter-service time $[2 \text{ min}, 5 \text{ min}]$ is equal to $46.5$\footnote{It is not clear how Terada came up with this value. If we exclude the times before the first southbound and northbound services and the times after the last southbound and northbound services from the set of inter-service times, the value should be replaced by $46 = 4.5 \times 3 + 3.5\times 5 + 2.5 \times 6$.}.
The sum of the product over the range $(0, 2 \text{ min}]$ is equal to $23.5$\footnote{This value also seems incorrect. The correct value should be $20.5 = 1.5 \times 10 + 0.5 \times 11$.}. The ratio of $46.5$ to $23.5$ is approximately equal to $2$ to $1$. If we incorporate the inter-service times larger than five minutes, then the ratio will be even more biased.

What does this mean?
Suppose that a passenger arrives at a station at a time drawn uniformly at random. Then, the probability that the passenger comes across an inter-service time of a specific length is equal to the product of the length of the inter-service time and the frequency of inter-service times with this length, divided by the sum of such products over all possible lengths of inter-service times. In our example, the passenger encounters an inter-service time shorter than two minutes once in three times, while the chance of encountering an inter-service time longer than two minutes and shorter than five minutes is twice in three times. In practice, we have to take into account the inter-service times longer than five minutes, so the latter probability would be arguably approximated by $3/4$. (We need some more detailed considerations to discuss the probability of the waiting time at a station, but I do not further it here.) This is just one example, but I have found similar tendencies in other observations that I made as well.

Anyways, passengers who always catch the first service to arrive without any consideration rarely come across a vacant service and are highly likely to get onto a congested one. Such experiences would be impressed on their memory.
Many occasions of traveling on congested services may probably be imprinted more strongly onto people's memory than rare occasions of traveling on vacant ones. If this is the case, the above proportional discrepancy\footnote{It is not very clear what exactly Terada means by ``proportional discrepancy." However, gathering from the context, it should be essentially the same thing as the waiting-time paradox.} will undergo a psychological transformation so that people may remember it in a somewhat self-amplified manner. Therefore, many people may forget the existence of vacant services and perceive that all services are congested.
This last issue on people's memory may be shaky. However, we have reached a solid conclusion, i.e., those who catch the first service to arrive are much more likely to encounter congested services than vacant ones. If more people get onto a congested service in this way, it will be even further delayed, eliciting further congestion.

This mechanism leads to odd conclusions. First, a great majority of passengers in Tokyo City trams get on a congested service by choice, although they are unconscious of it. Second, by doing so, those passengers are making efforts to enhance the congestion of these congested services.
These conclusions may appear to be paradoxical. However, this is an inevitable, rational consequence of the theory laid above. If this sounds odd, it is not because my argument is odd, but because the fact itself is odd.

If we want to avoid this situation, where the degree of congestion is heterogeneously distributed among trams, and want to secure a more balanced distribution, what we should do is obvious in theory. It is indeed necessary for the conductors or supervisors to disallow tram cars to be overloaded. An easier solution, however, is that the passengers themselves should hold back the desire to grab the first service to arrive to some extent, and seek a next vacant service even if they have to sacrifice 30 seconds to 2 minutes of their valuable time\footnote{Many commuters in Tokyo now still prefer 2 minutes to vacant trains. A century has not been long enough to change people's habit.}. I think that the loss of 30 seconds to 2 minutes would be reimbursed even before they reach the destination. 

However, whether or not one likes congested services is a personal matter. If many passengers have a particular interest in and enjoy competing with other passengers to get onto congested trams, it cannot be helped. A criterion with which one can discuss whether this taste is right or wrong is beyond mathematics or science.

In old days, giving way to others and sharing happiness with other people were counted as virtues. I do not know if this is the case now. However, putting aside this ethical question, even from an opportunistic or selfish point of view, at least for trams, it seems to be convenient and efficient behavior to give up congested services to other people, spend another minute, and then catch a vacant service, both for ourselves and others. It should be so at least for those who do not have a particular penchant to congestion.

The following is an aside --- I feel that we often encounter similar situations on our way in life.
It seems that there are two types of people, i.e., those who try hard to get on the first service to arrive and those who decide to wait a bit for the next service.
In our daily lives, things are so complicated that we have no idea how to apply simple mathematics to them. To what extent the analogy with the analysis of congested trams is relevant is beyond my imagination. Therefore, it is very difficult to judge which of these two types of behavior is good or bad.
This last problem should be too difficult for anybody. It may be a question of taste, which cannot be logically discussed. Here, I just want to make a point that there are various problems similar to the congested trams.
\end{mybox}
\setcounter{footnote}{\value{mpfootnote}}
\section{Further analysis of the waiting-time paradox in Tokyo City trams in 1922}
\label{sec:modern analysis}

It is amazing that an empirical study of the waiting-time paradox was done almost one century ago. This casual work could be claimed to belong to queuing theory, computational social science, or temporal network research in modern terms. It is a pity that this essay had not been available in English for a long time, at least
until a translation juxtaposed to the original Japanese text was published in a Japanese journal in 2013~\cite{Gally2013Kagaku-1,Gally2013Kagaku-2}. In this section, we re-assess what Terada did in more quantitative terms.

His scientific contributions are two-fold. First, he explicitly stated the waiting-time paradox and demonstrated it by analyzing inter-service times and waiting times that he observed in Tokyo City tramway services. Second, he casually discussed a positive feedback mechanism, whereby congested services would get more congested partly due to human behavior, and periodic dynamics of the level of congestion. Here we focus on the first result to further it. For the second effect, see~\cite{Nakatsuka1986JOperResSocJapan} for a mathematical analysis.

\begin{figure}[t]
    \centering
    \includegraphics[width=0.8\textwidth]{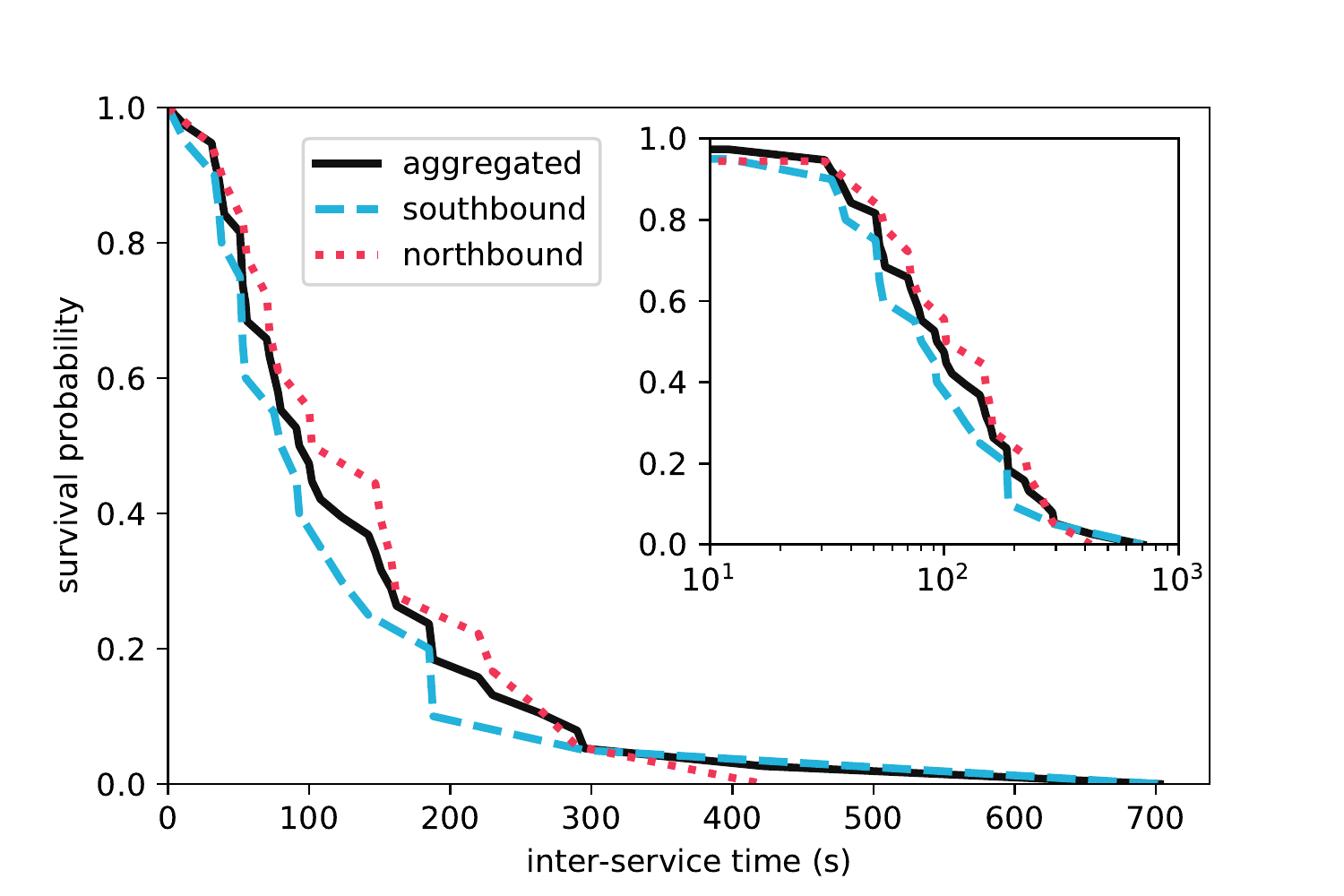}
    \caption{Survival probability functions of inter-service times for southbound trams (light blue dashed line), for northbound trams (red broken line), and for aggregated data (black solid line). The inset shows a semilog plot of the same survival probability functions. }
    \label{fig:IST_survival}
\end{figure}

We recalculated all the inter-service times based on the data that Terada provided in Fig.~\ref{fig:jimbocho_observation}.
The survival probability of inter-service times (i.e., the probability that the inter-service time is larger than a specified value), denoted by $\tau$, is shown in Fig.~\ref{fig:IST_survival}.
Here we did not use the Kaplan-Meier estimator, which is a more accurate means to estimate the survival probability of inter-event times~\cite{Kivela2015PhysRevE}. Furthermore, to take into consideration the large time interval before the first southbound service (see Fig.~\ref{fig:jimbocho_observation}), we counted the times from the beginning of the observation time window, i.e., 7:55 pm, to the first southbound and northbound services as two inter-service times. This is our arbitrary decision.
Although the sample size in Fig.~\ref{fig:IST_survival} is small, the figure suggests
a moderate level of dispersion in $\tau$ in both southbound and northbound services. The inset shows that the tail of the distribution of $\tau$ is roughly exponential, corresponding to Poisson processes.
The coefficient of variation (CV), defined by the standard deviation divided by the mean, of $\tau$
is shown in Table~\ref{tab:stats}. The table supports that the distribution of $\tau$ is roughly exponential; exponential distributions have a CV value of 1.
The mean inter-service time and the mean waiting time calculated by 
Eq.~\eqref{eq:mean waiting time} are also compared in Table~\ref{tab:stats}. We find that
the mean waiting time is roughly equal to the main inter-service time, $\langle \tau\rangle$, rather than $\langle \tau\rangle/2$. This result is quantitatively consistent with the effect of the waiting-time paradox for
Poisson processes explained in section~\ref{sec:waiting-time paradox}.

\begin{table}
\centering
\begin{tabular}{p{6em}p{4em}p{4em}p{4em}}
    & $\langle \tau \rangle$ & CV & $\langle t \rangle$\\ 
    \hline
    aggregated & 138.1 & 0.93 & 128.7\\
    southbound & 130.1 & 1.14 & 149.4\\
    northbound & 146.9 & 0.69 & 108.4\\
    \end{tabular}
    \caption{Mean inter-service time $\langle \tau \rangle$, CV of inter-service times, and mean waiting time $\langle t \rangle$. }
    \label{tab:stats}
\end{table}

\begin{figure}[t]
    \centering
    \includegraphics{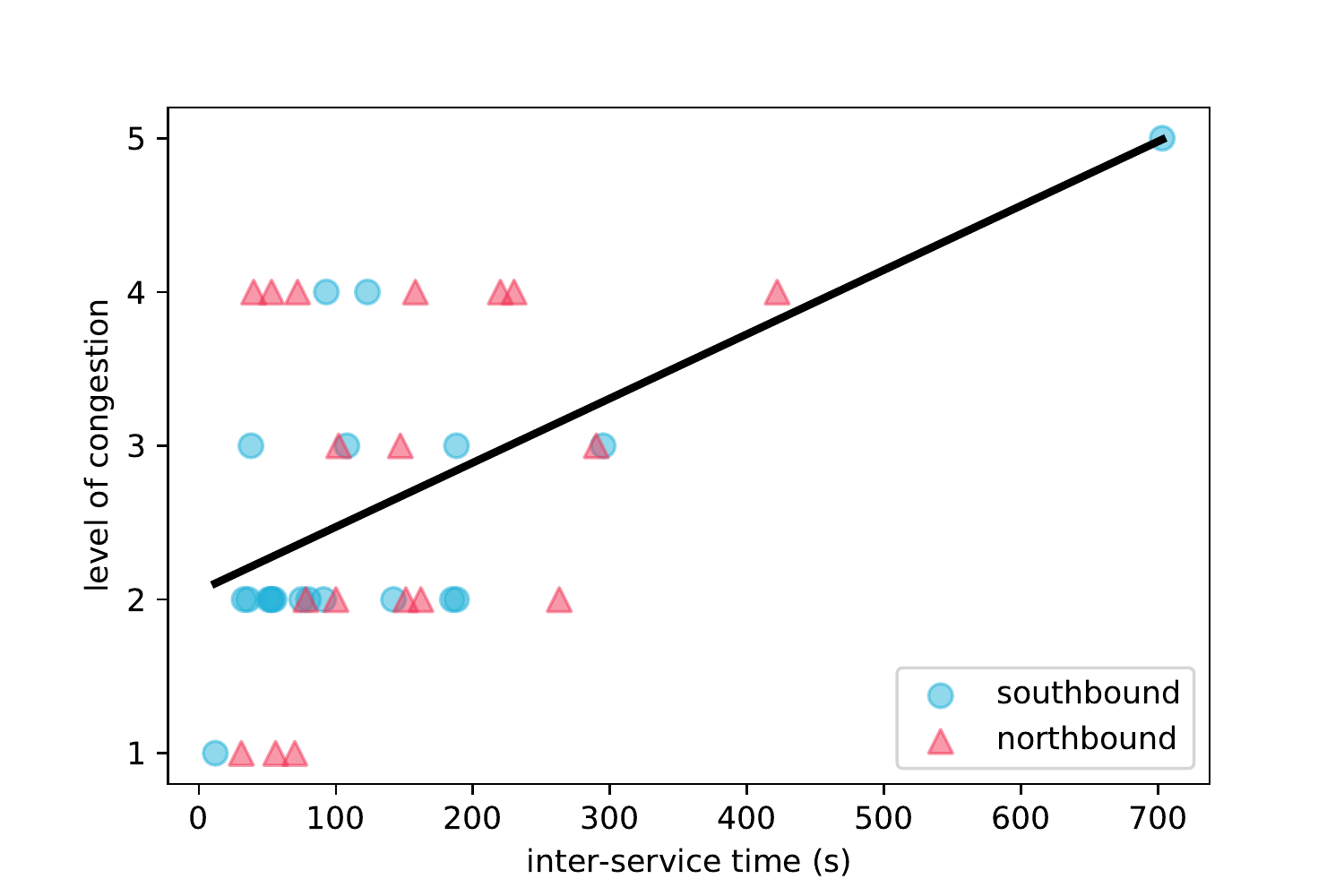}
    \caption{Correlation between the level of congestion of each service and the inter-service time preceding the service. The solid line represents a linear regression performed on the data for all services, $y = 0.0042\tau + 2.1$, where $y$ denotes the congestion level. 
We translated the five discrete levels of congestion explained in the caption of
Fig.~\ref{fig:jimbocho_observation} into $y\in \{1, 2, 3, 4, 5\}$.
If we exclude the first observations of southbound and northbound trams, then the data is fitted by regression $y = 0.0047\tau + 1.9$.}
    \label{fig:IST_crowdedness}
\end{figure}

The correlation between the level of congestion and the inter-service time is shown in Fig.~\ref{fig:IST_crowdedness}. We encoded the level of congestion according to Terada's casual observation into a five-point scale. The solid line shows the linear regression of the data. The Pearson correlation coefficient between the two variables is equal to 0.52, and significant ($p < 0.001$, $n=38$). Even if we exclude the time between 7:55 pm and the first observations of southbound and northbound services from the set of samples, the correlation remains significant ($p<0.01$, $n=36$). These results support Terada's casual observation that long inter-service times tend to be followed by congested services.

\section{Conclusions}
\label{sec:conclusion}

\begin{figure}
\centering
\includegraphics[width=0.35\linewidth]{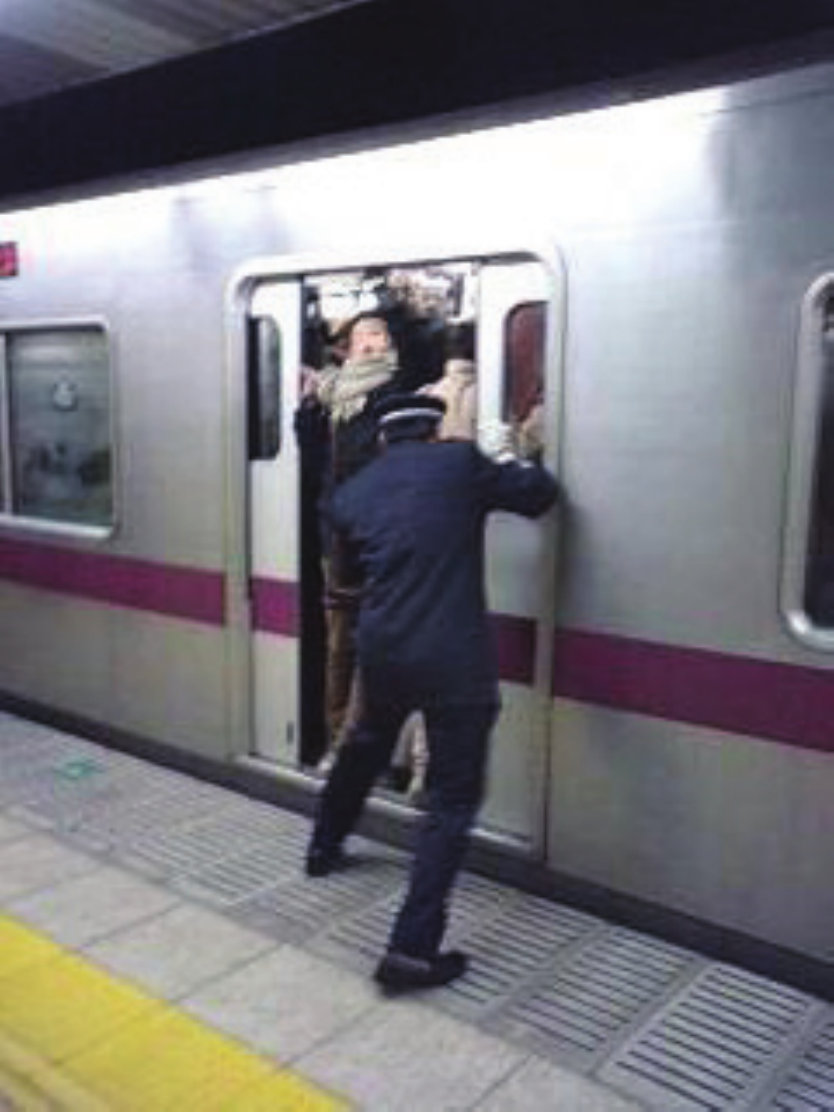}
\caption{Pushman in a train station in Tokyo.
The photo file is available on Japanese Wikipedia 
(\url{https://ja.wikipedia.org/wiki/\%E6\%8A\%BC\%E3\%81\%97\%E5\%B1\%8B}; Access date: April 18, 2020) and licensed under CC BY-SA 3.0, dated
07/02/2008 (Description: T\={o}ky\={u} Den-en-toshi line Sangenjaya station;
Source: Honya's file (self-made), Author: Honya; the English translation of the source and author names is by the present authors).}
\label{fig:pushman}
\end{figure}

We hope that readers enjoyed an old story about the waiting-time paradox from Japan.
Terada made observations of his and others' daily lives, formed a question based on them, constructed a theory using logic and mathematics, collected empirical data, and analyzed them to verify the theory. What he did was an exemplar scientific act. Perhaps to his surprise, the relevance of his observation and theorization has survived changes in the transportation systems and human behavior for a century. Congestion of trains in Tokyo now? See Fig.~\ref{fig:pushman}. As Terada stated, the level of congestion shown in this figure may be probably too high for his observation of oscillatory dynamics of congestion to apply. However, the waiting-time paradox is still there. Pushmen (see Fig.~\ref{fig:pushman}) are a modern invention to mitigate the waiting-time paradox as much as possible and to keep regular inter-service times. Passengers of Terada's type are still a minority in Tokyo.

\section*{Acknowledgments}
We acknowledge Aozora Bunko project (\url{http://www.aozora.gr.jp/}; only in Japanese), which digitized Ref.~\cite{Terada1922essay} and generously permits translation and modification. 
According to the Copyright Act of Japan, the copyright of the original article has expired because the author deceased more than 50 years ago. 


\begin{thebibliography}{10}

\bibitem{Terada1913Nature-1}
T.~Terada.
\newblock {X-rays and crystals}.
\newblock {\em Nature}, 91:135--136, 1913.

\bibitem{Terada1913Nature-2}
T.~Terada.
\newblock {X-rays and crystals}.
\newblock {\em Nature}, 91:213, 1913.

\bibitem{Terada1913ProcMathPhysSocTokyo}
T.~Terada.
\newblock {On the transmission of X-rays through crystals}.
\newblock {\em Proc. Math. Phys. Soc. Tokyo}, 7:60--70, 1913.

\bibitem{Matsushita2010EvolInstEconRev}
M.~Matsushita.
\newblock {Torahiko Terada (1878–1935): Father of the science of complex
  systems}.
\newblock {\em Evol. Inst. Econ. Rev.}, 6:337--340, 2010.

\bibitem{Wittner2016book}
D.~G. Wittner and P.~C. Brown, editors.
\newblock {\em Science, Technology, and Medicine in the Modern Japanese
  Empire}.
\newblock Routledge, Abingdon, UK, 2016.

\bibitem{Nakatsuka1986JOperResSocJapan}
T.~Nakatsuka.
\newblock {Periodic property of streetcar congestion at the first station}.
\newblock {\em J. Oper. Res. Soc. Japan}, 29:1--20, 1986.

\bibitem{Gally2013Kagaku-1}
T.~Gally and M.~Matsushita.
\newblock {Torahiko in English (11)}.
\newblock {\em Kagaku}, 83:274--281, Mar. 2013.

\bibitem{Gally2013Kagaku-2}
T.~Gally and M.~Matsushita.
\newblock {Torahiko in English (12)}.
\newblock {\em Kagaku}, 83:574--582, May 2013.

\bibitem{Feller1971book2}
W.~Feller.
\newblock {\em An Introduction to Probability Theory and Its Applications,
  Volume II, Second Edition}.
\newblock John Wiley \& Sons, New York, 1971.

\bibitem{Allen1990book}
A.~O. Allen.
\newblock {\em Probability, Statistics, and Queueing Theory: With Computer
  Science Applications, Second Edition}.
\newblock Academic Press, Boston, 1990.

\bibitem{Masuda2016book}
N.~Masuda and R.~Lambiotte.
\newblock {\em A Guide to Temporal Networks}.
\newblock World Scientific, Singapore, 2016.

\bibitem{Eckmann2004PNAS}
J.~P. Eckmann, E.~Moses, and D.~Sergi.
\newblock {Entropy of dialogues creates coherent structures in e-mail traffic}.
\newblock {\em Proc. Natl. Acad. Sci. USA}, 101:{14333}--{14337}, 2004.

\bibitem{Barabasi2005Nature}
A.~L. Barab\'{a}si.
\newblock {The origin of bursts and heavy tails in human dynamics}.
\newblock {\em Nature}, 435:207--211, 2005.

\bibitem{VazquezA2006PhysRevE-burst}
A.~V\'{a}zquez, J.~G. Oliveira, Z.~Dezs\"{o}, K.~I. Goh, I.~Kondor, and A.~L.
  Barab\'{a}si.
\newblock {Modeling bursts and heavy tails in human dynamics}.
\newblock {\em Phys. Rev. E}, 73:{036127}, 2006.

\bibitem{GohBarabasi2008EPL}
K.~I. Goh and A.~L. Barab\'{a}si.
\newblock {Burstiness and memory in complex systems}.
\newblock {\em Europhys. Lett.}, 81:48002, 2008.

\bibitem{Barabasi2010book}
A.~L. Barab\'{a}si.
\newblock {\em Bursts --- The Hidden Pattern behind Everything We Do}.
\newblock Dutton, New York, 2010.

\bibitem{HolmeSaramaki2012PhysRep}
P.~Holme and J.~Saram\"{a}ki.
\newblock {Temporal networks}.
\newblock {\em Phys. Rep.}, 519:97--125, 2012.

\bibitem{Karsai2018Springer}
M.~Karsai, H.~H. Jo, and K.~Kaski.
\newblock {\em Bursty Human Dynamics}.
\newblock Springer, Cham, Switzerland, 2018.

\bibitem{Masuda2013F1000}
N.~Masuda and P.~Holme.
\newblock {Predicting and controlling infectious disease epidemics using
  temporal networks}.
\newblock {\em F1000Prime Reports}, 5:6, 2013.

\bibitem{Terada1922essay}
T.~Terada.
\newblock Densha no konzatsu ni tsuite [{On congested trams}].
\newblock {\em Shiso}, 12:45--57, Sep. 1922.

\bibitem{Kivela2015PhysRevE}
M.~Kivel\"{a} and M.~A. Porter.
\newblock {Estimating interevent time distributions from finite observation
  periods in communication networks}.
\newblock {\em Phys. Rev. E}, 92:052813, 2015.

\end{thebibliography}

\end{document}